\documentclass[aps,prd,superscriptaddress,tightenlines, nofootinbib,preprintnumbers,floatfix,12pt]{revtex4-2}

\usepackage{graphicx}
\usepackage{multirow}
\usepackage{enumitem}
\usepackage{amssymb}
\usepackage{amsmath}
\usepackage{xcolor}
\usepackage{xspace,ulem}

\usepackage[colorlinks=true,citecolor=blue,urlcolor=blue,linkcolor=blue]{hyperref}

\newcommand{\nn}{\nonumber}
\newcommand{\beq}{\begin{equation}}
\newcommand{\eeq}{\end{equation}}
\newcommand{\beqa}{\begin{eqnarray}}
\newcommand{\eeqa}{\end{eqnarray}}

\newcommand{\znbb}{\ensuremath{0\nu\beta\beta}\xspace}
\newcommand{\mee}{\ensuremath{m_{ee}}\xspace}
\newcommand{\mmue}{\ensuremath{m_{\mu e}}\xspace}

\newcommand{\malphabeta}{\ensuremath{m_{\alpha\beta}}\xspace}

\def\OMIT#1{{}}

\makeatletter
\g@addto@macro\bfseries{\boldmath}
\let\Hy@backout\@gobble
\makeatother

\newcommand{\nuMSM}{$\nu_M{\rm SM}$\xspace}
\newcommand{\nuDSM}{$\nu_D{\rm SM}$\xspace}

\begin{document}

\title{Majorana phases beyond neutrinoless double beta decay}

\author{Avital Dery}
\affiliation{Department of Physics, LEPP, Cornell University, Ithaca, NY 14853, USA}

\author{Stefania Gori}
\affiliation{Department of Physics and Santa Cruz Institute for Particle Physics, University of California Santa Cruz, Santa Cruz, CA 95064, USA}

\author{Yuval Grossman}
\affiliation{Department of Physics, LEPP, Cornell University, Ithaca, NY 14853, USA}

\author{Zoltan Ligeti}
\affiliation{Ernest Orlando Lawrence Berkeley National Laboratory, 
University of California, Berkeley, CA 94720, USA}

\begin{abstract}

The \nuMSM is defined as the SM extended to include dimension-5 operators. 
In this model neutrino masses violate lepton number, and two parameters of the lepton mixing matrix, the Majorana phases, are yet to be constrained. 
One combination of these phases and the neutrino masses, often denoted by \mee, is probed by neutrinoless double beta decays (\znbb).
We explore what information may be obtained beyond \znbb, and how it depends on the lightest neutrino mass. 
We point out that with current central values of the mixing parameters, $\Delta L_e=2$ and $\Delta L_e=\Delta L_\mu = 1$ 
(or $\Delta L_e=2$ and $\Delta L_\mu = 2$) 
processes cannot simultaneously vanish, providing a no-lose theorem, in principle, for 
excluding the \nuMSM, even in the case of normal mass ordering.

\end{abstract}

\maketitle

%%%%%%%%%%%%%%%%%%%%%%%%%%%%%%%%%%%%%
%%%%%%%%%%%%%%%%%%%%%%%%%%%%%%%%%%%%%
\section{Introduction}

The fact that neutrinos are massive implies that the Standard Model (SM) must be extended. One option is extending the SM to include dimension-5 operators, without introducing new fields, yielding lepton number violating (LNV) Majorana mass terms. We refer to this scenario as the \nuMSM.
Another option is to add right-handed singlet fields to the SM and impose lepton number conservation, or assume that only Dirac mass terms arise for the neutrinos for some other reason. 
We denote this model as the \nuDSM. 

The number of parameters in the PMNS~\cite{Maki:1962mu,Pontecorvo:1967fh} lepton mixing matrix depends on which option is realized in Nature.
In the \nuDSM scenario, the PMNS matrix is parametrized, just like the CKM matrix, by three mixing angles and a single $CP$ violating phase, usually called the ``Dirac phase." 
In the \nuMSM scenario, there are two additional physical phases, which can only affect LNV processes. These two extra phases are usually referred to as ``Majorana phases."

The PMNS matrix is commonly parameterized as~\cite{PDG24}
\begin{align}\label{pmns}
U =& \begin{pmatrix} 1 & & \\ & c_{23} & s_{23} \\ & -s_{23} & c_{23} \end{pmatrix} \times
  \begin{pmatrix} c_{13} &  & s_{13}\, e^{-i\delta} \\ & 1 & \\ -s_{13}\, e^{i\delta} &  & c_{13} \end{pmatrix} 
 \times \begin{pmatrix} c_{12} & s_{12} & \\ -s_{12} & c_{12} & \\ & & 1 \end{pmatrix} \times
  \begin{pmatrix} e^{i\eta_1} & & \\  & e^{i\eta_2} & \\ & & 1 \end{pmatrix} ,
\end{align}
where $c_{ij} = \cos\theta_{ij}$, $s_{ij} = \sin\theta_{ij}$, $ij = \{12,\, 23,\, 13\}$, with $\theta_{ij}$ the mixing angles, $\delta$ is the $CP$ violating Dirac phase, and $\eta_{1,2}$ are the Majorana phases. The Majorana phases are not present in the \nuDSM.

There is significant experimental knowledge on the values of neutrino masses and the PMNS parameters, as summarized in Table~\ref{tab:values}. There are two options for the ordering of the masses: 
normal ordering (NO), where the smaller mass difference is between the two lighter states and inverted ordering (IO), where the smaller mass difference is between the two heavier states.
If NO is established in the future, it would imply that \znbb might never be observed, and exploring other possibilities to detect lepton number violation becomes even more important.
While so far we do not have much information on the Dirac phase, $\delta$, it is expected to be measured by Hyper-Kamiokande~\cite{Hyper-Kamiokande:2022smq} and DUNE~\cite{DUNE:2022aul}. 

\begin{table}[t]
\tabcolsep 6pt
\begin{tabular}{c|cc}
\hline\hline
\multirow{2}{*}{Parameter}  &  \multicolumn{2}{c}{Value} \\
  &  NO  &  IO \\
\hline 
$\theta_{12}$  &  $\big(33.41^{+0.75}_{-0.72}\big)^\circ$  &  same \\
$\theta_{23}$  &  $\big(49.1^{+1.0}_{-1.3}\big)^\circ$  &  $\big(49.5^{+0.9}_{-1.2}\big)^\circ$\\
$\theta_{13}$  &  $\big(8.54 ^{+0.11}_{-0.12}\big)^\circ$  &  $\big(8.57^{+0.12}_{-0.11}\big)^\circ$\\
${\Delta m_{21}^2}/({10^{-5}\,{\rm eV}^2})$ &  $7.41^{+0.21}_{-0.20}$ &  same\\
${\Delta m_{32}^2}/({10^{-3}\,{\rm eV}^2})$ & $2.437^{+0.028}_{-0.027}$ & $-2.498^{+0.032}_{-0.025}$\\
\hline\hline
\end{tabular}
\caption{Values of the mixing angles and mass squared differences used in our analysis.  We use the ``w/o SK-ATM" fit results in the PDG review~\cite{PDG24}. 
}
\label{tab:values}
\end{table}

There are several ways to constrain $\sum m_i$. Laboratory experiments give bounds of the order of 1\,eV~\cite{PDG24, Katrin:2024tvg}.  Cosmology gives stronger bounds.  Assuming minimal $\Lambda$CDM, which is not a very good description of the data, the DESI fit to BAO + CMB measurements suggests $\sum m_i < 0.072\,$eV (95\% CL)~\cite{DESI:2024mwx}.  At the same time, assuming a lower bound on $\sum m_i$ as a prior in the DESI analysis, the preference for NO is only mild. 
We view the mass ordering as an open question and consider values of the lightest neutrino mass up to $0.1$ eV.

In principle, one can distinguish between the \nuDSM and the \nuMSM experimentally.  Observing any lepton number violating process would rule out the \nuDSM. The most promising probes are neutrinoless double beta decay (\znbb) experiments, measuring the following combination of masses and PMNS parameters,
\beq\label{meedef}
\mee = \bigg| \sum^3_{i=1}\, m_i\, U^2_{ei} \bigg| \,,
\eeq
where $m_i$ are the three physical neutrino masses. 
The situation is more complicated in trying to rule out the \nuMSM.  It can be excluded in the case of IO, by not observing \znbb.
However, in case of NO, the \znbb rate may vanish even if neutrinos are Majorana.
In the \nuMSM and for IO, there is a lower bound from our knowledge of mixing angles and mass-squared differences, $m_{ee} > 16\,{\rm meV}$ (at 95\%\;CL)~\cite{PDG24}.
Using a variety of nuclear matrix element calculations, the current experimental upper bound from \znbb searches is $m_{ee} < (22 - 122)\,{\rm meV}$~\cite{KamLAND-Zen:2024eml}.
On the other hand, in the case of
NO, for certain values of the lightest neutrino mass, the \znbb rate may be arbitrarily small, even if the neutrino mass terms violate lepton number. 
In this case, a null result at \znbb experiments will not determine the nature of neutrino masses.

If \znbb is observed, one combination of the two Majorana phases will be measured. 
This combination depends on the neutrino masses, as given in Eq.~(\ref{meedef}).
Other LNV processes with two charged leptons are, in principle, sensitive to other combinations of the Majorana phases, although the prospects for their detection within the \nuMSM currently seem remote.
It is convenient to define the generalized effective Majorana mass matrix,
\beq\label{mabdef}
\malphabeta = \bigg| \sum^3_{i=1}\, m_i\, U_{\alpha i} U_{\beta i} \bigg|\,, \qquad \alpha,\beta \in \{e,\mu,\tau\}\,. 
\eeq

The properties of LNV processes and their dependence on the Majorana phases have been discussed in the literature~\cite{Xing:2013woa,Simkovic:2012hq,Barger:2002vy,Pascoli:2002qm,Pascoli:2005zb,deGouvea:2002gf,Flanz:1999ah,Rodejohann:2000hy,Atre:2005eb,Abada:2017jjx,Ali:2001gsa,Bar-Shalom:2006osy,Rodejohann:2000ne,Rodejohann:2000th,Zuber:2000vy}. 
The prospects for constraining one combination of Majorana phases from \znbb and for establishing $CP$ violation have been studied in Refs.~\cite{Simkovic:2012hq,Barger:2002vy,Pascoli:2002qm,Pascoli:2005zb}. 
Ref.~\cite{deGouvea:2002gf} studied the role of Majorana phases in manifestly $CP$ violating effects in the lepton sector. 
Refs.~\cite{Flanz:1999ah,Rodejohann:2000hy,Atre:2005eb} considered the full effective Majorana mass matrix, \malphabeta, and the experimental bounds on all six elements, and Ref.~\cite{Abada:2017jjx} studied \malphabeta in the context of models with extra GeV-scale sterile states. Refs.~\cite{Ali:2001gsa,Bar-Shalom:2006osy} studied collider signatures of $\Delta L=2$ processes. 

While there are ongoing searches for \znbb, and thus for probing $m_{ee}$, 
the prospects of observing other LNV processes are less promising.
In the near future, the best sensitivity to a lepton number violating process involving not only electrons will probably come from $\mu^-$ to $e^+$ conversion.  The Mu2e~\cite{Mu2e-II:2022blh} and COMET~\cite{COMET:2018auw} experiments will search for the $pp\mu^- \to nne^+$ rate.
Mu2e expects to be sensitive at the $10^{-16}$ level~\cite{Diociaiuti:2020fjx}, many orders of magnitudes away from the much lower rates that could arise in the \nuMSM , of order $(10^{-40})$~\cite{Domin:2004tk,Lee:2021hnx}, which would be proportional to $m_{\mu e}^2$.
%\ZL{I'd keep the part you suggest to remove, otherwise this paragraph never mentions that $\mu^-$ to $e^+$ conversion would probe $m_{\mu e}$.}\AD{Agreed.}

Throughout this paper we treat LNV processes as observables of interest, even though the prospects of measuring any LNV process within the \nuMSM besides \znbb seem slim at the moment. 
Therefore, our results are of a theoretical nature, though one may hope to find new ways to probe these quantities in the future.
In the following, we interpret LNV observables within the \nuMSM. We use the parameter values and uncertainties shown in Table~\ref{tab:values}. 
When $\delta$ plays a role in our analysis, we either vary it in the full range or use the projections by Hyper-Kamiokande, where the uncertainty of $\delta$ is expected to reach $7^\circ$ ($22^\circ$) for $\delta=0$ ($90^\circ$) after 10\,years of data taking~\cite{Hyper-Kamiokande:2022smq}.

We first discuss the physically observable combinations of phases using rephasing invariants defined in Section~\ref{sec:physicalPhases}. In Section~\ref{sec:general} we point out several general properties of the effective Majorana mass matrix of Eq.~\eqref{mabdef}. In Section~\ref{sec:singlePhase} we derive a simplified dependence on Majorana phases in the regime of a hierarchical mass spectrum. 
The intriguing possibility to arrive at a theoretical no-lose theorem for the Majorana nature of neutrino masses using the combination of \mee and \mmue is discussed in Section~\ref{sec:nolose}. We conclude in Section~\ref{sec:conclusion}.

%%%%%%%%%%%%%%%%%%%%%%%%%%%%%%%%%%
%%%%%%%%%%%%%%%%%%%%%%%%%%%%%%%%%%
\section{Physical Dirac and Majorana phases}
\label{sec:physicalPhases}

The distinction between Dirac and Majorana phases is subtle. 
The phase parameters in Eq.~\eqref{pmns} are convention dependent (for example, the Majorana phases $\eta_1$ and $\eta_2$ can be shifted by a global rephasing of $U$), and therefore cannot correspond to physical quantities. 
Instead, we define Dirac and Majorana phases based on how they affect observables.
We define a Dirac phase as a $CP$ violating phase that is measured in lepton number conserving processes, such as neutrino oscillations.
Majorana phases are those which are only accessible through LNV processes.

For this purpose, it is useful to work with quantities that are rephasing invariant. 
We introduce the following phase-convention independent quantities to describe the physical phases~\cite{Nieves:1987pp},
\begin{align}\label{invariants}
    t_{\alpha i \beta j} \, = \, U_{\alpha i}U_{\beta j}U_{\alpha j}^* U_{\beta i}^* \,, \qquad
    s_{\alpha ij} \, = \, U_{\alpha i}U_{\alpha j}^* \,.
\end{align}
We refer to $t_{\alpha i \beta j}$ as
quartic invariants  and to $s_{\alpha ij}$ as quadratic invariants. Note that both types of invariants are hermitian with respect to the mass eigenstate indices, $\{i, j\}$. The $t_{\alpha i \beta j}$ are also hermitian with respect to the flavor indices, $\{\alpha,\beta\}$.
We denote the phases as
\begin{equation}
\Phi^{\alpha}_{ij}\equiv \arg(s_{\alpha i j})\,, \qquad
\label{eq:deltaalphabetaij}
\delta^{\alpha\beta}_{ij} \equiv \arg(t_{\alpha i \beta j})\, .
\end{equation}

We define two transformations
\beq
T_m:= U\to U\, {\rm diag}\big(e^{i\phi_1}, e^{i\phi_2}, e^{i\phi_3}\big)\,, \qquad
T_f:=U\to {\rm diag}\big(e^{i\phi_e}, e^{i\phi_\mu},e^{i\phi_\tau}\big)\, U\,,
\eeq
where the indices $m$ and $f$ indicate mass and flavor.
The $t_{\alpha i \beta j}$ are invariant under both $T_m$ and $T_f$.
The $s_{\alpha ij}$ are invariant under $T_f$ but they are not invariant under $T_m$ transformations.

A basis for the $t_{\alpha i \beta j}$ invariants consists of three magnitudes and a phase, and can be chosen as 
\begin{equation}\label{eq:tbasis}
	\big\{ |t_{e2e3}|,\ |t_{e3e3}|,\ |t_{\mu2 e3}|,\ \Psi_D \big\}\, ,
\end{equation}
where we define an invariant Dirac phase as 
\beq\label{phiDdef}
\Psi_D \equiv \delta^{\mu e}_{23} = \arg(t_{\mu 2 e 3}) \, .
\eeq
All quartic invariants can be expressed in terms of the four quantities in Eq.~(\ref{eq:tbasis}), using the following relations,
\begin{align}
t_{\alpha i \beta j} &= t_{\alpha i \beta k}\, \frac{t_{\alpha j \beta k}^*}{t_{\alpha k \beta k}}\,, \qquad\qquad
  t_{\alpha i \beta j} = t_{\alpha i \gamma j}\, \frac{t_{\beta i \gamma j}^*}{t_{\gamma i \gamma j}} \,, \nn\\
\sum_i t_{\alpha i \beta j} &= \delta_{\alpha\beta}\, \sqrt{t_{\alpha j \alpha j}}\,,  \qquad\, 
  \sum_\alpha t_{\alpha i \beta j} = \delta_{ij}\, \sqrt{t_{\beta i \beta j}}\,.
\end{align}
We note that the area of the unitarity triangles in the lepton sector can be expressed as $J = s_{12}s_{23}s_{13}c_{12}c_{23}c^2_{13} \sin\delta = -|t_{\mu 2 e 3}|\sin\Psi_D$.

The four parameters in Eq.~\eqref{eq:tbasis} suffice to describe the PMNS matrix in the \nuDSM.
For the case of the \nuMSM, the additional phase convention independent parameters needed to describe Majorana phases arise from the $s_{\alpha ij}$ invariants, which are physical in this case since the $T_m$ transformation is no longer a symmetry in this model.

Of the nine non-vanishing $\Phi^{\alpha}_{ij}$ phases, two are independent.
It is convenient to first eliminate all $\Phi^{\alpha}_{13}$'s by using the identity
\begin{equation}\label{eq:iden1}
	\Phi^{\alpha}_{13} = \Phi^{\alpha}_{12} + \Phi^{\alpha}_{23}\, . 
\end{equation}
Of the remaining six phases, only one of the three, $\{\Phi^{e}_{12},\, \Phi^{\mu}_{12},\, \Phi^{\tau}_{12}\}$,
and one of the three, $\{\Phi^{e}_{ 23},\, \Phi^{\mu}_{23},\, \Phi^{\tau}_ {23}\}$,
are independent.
The rest can be expressed in terms of the chosen two and the Dirac phase, using a second identity,
\begin{equation}\label{eq:iden2}
	\Phi^{\beta}_{ij} = \Phi^{\alpha}_{ij} + \delta^{\beta\alpha}_{ij}\,.
\end{equation}

Therefore, a basis for the Majorana phases could be any pair, 
\begin{equation}
    \{\Phi^{\alpha}_{12},\, \Phi^{\beta}_{23}\}\, \qquad \alpha,\beta = e,\mu,\tau\,  ,
\end{equation}
or any pair of linear combinations, $\big\{ \sum_\alpha a_\alpha\, \Phi^{\alpha}_{12}, \ \sum_\beta b_\beta\, \Phi^{\beta}_{23} \big\}$.
Throughout this paper we primarily use the following choice of basis, which is convenient for \znbb,
\begin{equation}\label{eq:Mphasedef}
    \{\Phi_{12},\, \Phi_{23}\} \, \equiv \, \{\Phi^{e}_{12},\, \Phi^{e}_{23}\}\, .
\end{equation}
In terms of the convention for the PMNS matrix in Eq.~\eqref{pmns}, the physical Dirac and Majorana phases, as defined in Eqs.~(\ref{phiDdef}) and (\ref{eq:Mphasedef}), are given by
\begin{align}\label{eq:3phaseRelation}
    \Psi_D &= \arg\big(c_{12}c_{23}e^{-i\delta}-s_{12}s_{23}s_{13}\big)\, ,  \nn\\
	\Phi_{12} &= \eta_1-\eta_2\, ,  \nn\\
    \Phi_{23} & = \eta_2+\delta \, .
\end{align}

%%%%%%%%%%%%%%%%%%%%%%%%%%%%%%%%%%%%%%%%%%%%%%
%%%%%%%%%%%%%%%%%%%%%%%%%%%%%%%%%%%%%%%%%%%%%%
\section{General statements}
\label{sec:general}

Within the framework of the \nuMSM, the following three statements hold:
\begin{enumerate}
\item 
The rate of any LNV process involving two charged leptons of flavor $\alpha$ and $\beta$ is proportional to 
$\malphabeta^2$ where $\malphabeta$ is defined in Eq.~(\ref{mabdef}).
\end{enumerate}
In the \nuMSM, in the absence of right-handed charged current interactions, the amplitude of any LNV process with two charged leptons takes the following form, 
\beq
i {\cal M}^{\mu\nu} \propto \sum_i\bar u \gamma^\mu P_L U_{\alpha i}\, \frac{i(p_\rho \gamma^\rho+m_i)C}{p^2-m^2}\, U_{\beta i}(\bar u \gamma^\nu P_L)^T \, \propto m_{\alpha\beta}  \,,
\eeq
with $C=-i\gamma^2\gamma^0$.
Due to the Dirac structure, only the part of the neutrino propagator proportional to the mass contributes,
which is what leads to $\malphabeta$.

\begin{enumerate}\setcounter{enumi}{1}
\item 
Any one element of the effective Majorana mass matrix, $\malphabeta$, is in itself independent of the Dirac phase.
This implies that prior knowledge of $\Psi_D$ is not required for a prediction of any element, $\malphabeta$.
In particular, knowledge of the Dirac phase does not affect the uncertainty of the prediction for $\znbb$. 
\end{enumerate}
To demonstrate this second point, we write the square of the effective Majorana mass as
\begin{eqnarray}\label{eq:mabsquare}
\malphabeta^2 \,&=\,   \sum_{i,j} m_i m_j s_{\alpha ij}s_{\beta ij} \\ \nonumber
&= \sum_i m_i^2 |U_{\alpha i}|^2|U_{\beta i}|^2  \, &+ \,2m_1 m_2 |U_{\alpha 1}U_{\alpha 2}U_{\beta 1}U_{\beta 2}|\cos\big(\Phi^{\alpha}_{12}+\Phi^{\beta}_{12}\big)	 \\ \nonumber
&	& + \,2m_2 m_3 |U_{\alpha 2}U_{\alpha 3}U_{\beta 2}U_{\beta 3}|\cos\big(\Phi^{\alpha}_{23}+\Phi^{\beta}_{23}\big)	 \\ \nonumber
&	& + \,2m_1 m_3|U_{\alpha 1}U_{\alpha 3}U_{\beta 1}U_{\beta 3}|\cos\big((\Phi^{\alpha}_{12}+\Phi^{\beta}_{12})+(\Phi^{\alpha}_{23}+\Phi^{\beta}_{23})\big)\,,
\end{eqnarray}
where for the last term we have used Eq.~(\ref{eq:iden1}).
We see that for any given $m_{\alpha\beta}$, we can choose a basis such that no Dirac phase appears. That is, if we use the basis
\begin{eqnarray}\label{eq:alphabetabasis}
\Phi^{\alpha\beta}_{12} = \frac{\Phi^{\alpha}_{12} + \Phi^{\beta}_{12}}{2} \,, \qquad
\Phi^{\alpha\beta}_{23} = \frac{\Phi^{\alpha}_{23} + \Phi^{\beta}_{23}}{2}\, ,
\end{eqnarray}
then Eq.~\eqref{eq:mabsquare} has the form
\begin{eqnarray}\label{eq:master}
	m_{\alpha\beta}^2  &=  \sum_i m_i^2 |U_{\alpha i}|^2|U_{\beta i}|^2  \, &+ \,2m_1 m_2 |U_{\alpha 1}U_{\alpha 2}U_{\beta 1}U_{\beta 2}|\cos\big(2\Phi^{\alpha\beta}_{12}\big)	 \\ \nonumber
	&  &+ \,2m_2 m_3 |U_{\alpha 2}U_{\alpha 3}U_{\beta 2}U_{\beta 3}|\cos\big(2\Phi^{\alpha\beta}_{23}\big)	 \\ \nonumber
	& &+ \,2m_1 m_3|U_{\alpha 1}U_{\alpha 3}U_{\beta 1}U_{\beta 3}|\cos\big(2\Phi^{\alpha\beta}_{12}+2\Phi^{\alpha\beta}_{23}\big)\, .
\end{eqnarray}
From Eq.~\eqref{eq:master} it is clear that, for any $\{\alpha,\beta\}$, $m_{\alpha\beta}^2$ depends only on two phases, which can be expressed as pure Majorana phases.
This is true for any one $m_{\alpha\beta}$ observable alone. As soon as more than one is considered, there is no choice of basis for the Majorana phases that allows both to be expressed without the use of Dirac phases. 
Thus, while $m_{\alpha\beta}$ can be expressed as a function of the two Majorana phases in Eq.~\eqref{eq:alphabetabasis},
considering a different $m_{\alpha^\prime \beta^\prime}$, expressed in the same basis, the phases in the cosines would also include Dirac phases,
\begin{eqnarray}
    \Phi^{\alpha^\prime\beta^\prime}_{12} &=& \Phi^{\alpha\beta}_{12} + (\delta^{\alpha^\prime \alpha}_{12}+ \delta^{\beta^\prime \beta}_{12})/2\, ,\\ \nonumber
    \Phi^{\alpha^\prime\beta^\prime}_{12} &=& \Phi^{\alpha\beta}_{23} +  (\delta^{\alpha^\prime \alpha}_{23}+ \delta^{\beta^\prime \beta}_{23})/2 \, ,
\end{eqnarray}
where we have used the identity in Eq.~\eqref{eq:iden2}.
We see that when considering more than one LNV observables, all three phases play a role. In principle, measurement of three different $m_{\alpha\beta}$ could determine all three phases --- two Majorana and one Dirac.

The lack of dependence on the Dirac phase can be obscured when working in a specific convention, such as that of Eq.~\eqref{pmns}. 
For example, $\mee^2$ in the basis of Eq.~\eqref{pmns} reads,
\begin{align}\label{eq:meeSqrpmnsbasis}
    \mee^2 \, &= \, (m_1^2c_{12}^4 + m_2^2 s_{12}^4) c_{13}^4 + m_3^2s_{13}^4 +2m_1m_2 s_{12}^2c_{12}^2 c_{13}^4 \cos\big[2(\eta_1-\eta_2)\big] \nn\\
    & \quad + 2m_1 m_3 c_{12}^2s_{13}^2c_{13}^2\cos\big[2(\eta_1+\delta)\big] + 2m_2m_3s_{12}^2s_{13}^2c_{13}^2\cos\big[2(\eta_2+\delta)\big]\, .
\end{align}
While this expression depends on $\delta$, varying the value of $\delta$ has no effect on the possible values the three cosine terms can take.
This is because $\delta$ only enters as $(\eta_1+\delta)$ and $(\eta_2+\delta)$, and thus it simply shifts the extracted values of $\eta_{1,2}$.
Therefore, the predictions for $\mee$ with unknown Majorana phases are not affected by any knowledge of the Dirac phase (or lack thereof). 
The same holds when considering any one element of the \malphabeta matrix by itself.  However, independence of the Dirac phase, $\Psi_D$, does not necessarily imply independence of the parameter $\delta$, which in the parametrization of Eq.~(\ref{pmns}) also affects magnitudes of some PMNS matrix elements.
\begin{enumerate}\setcounter{enumi}{2}
\item 
Sensitivity to each of the Majorana phases, $\Phi_{ij}$, scales as the corresponding product of masses, $m_i\, m_j$.
\end{enumerate}
Rewriting Eq.~\eqref{eq:master} using the Majorana phase basis of Eq.~\eqref{eq:Mphasedef} ($\Phi_{ij}\equiv \Phi^{e}_{ij}$), we have
\begin{equation}\label{eq:masteree}
	\malphabeta^2 \,=\,  \sum_i m_i^2\, |U_{\alpha i}|^2 |U_{\beta i}|^2  
     +  2\sum_{i<j} m_i m_j\, \big|U_{\alpha i} U_{\alpha j}U_{\beta i}U_{\beta j}\big| \cos\big(2\Phi_{ij} + \delta^{\alpha e}_{ij} + \delta^{\beta e}_{ij}\big)\, .
\end{equation}
Eq.~\eqref{eq:masteree} shows that a given Majorana phase, $\Phi_{ij}$, accompanies the corresponding product of masses, $m_i\, m_j$, in the expression for $\malphabeta^2$, regardless of the flavor indices $\alpha,\,\beta$. 
This is not a surprise, since the Majorana phases are those that could have been rotated away via rephasing of the neutrino mass eigenstates, were the mass terms of the Dirac type.

%%%%%%%%%%%%%%%%%%%%%%%%%%%%%%%%%%%%%%%%%%%%%%%%%%%%%
%%%%%%%%%%%%%%%%%%%%%%%%%%%%%%%%%%%%%%%%%%%%%%%%%%%%%
\section{Approximate relations for hierarchical masses}
\label{sec:singlePhase}

In the following we consider the limit where the lightest neutrino is very light (we discuss below the numerical values of when this limit is justified). 
In this case we point out that, to leading order, all $m_{\alpha\beta}$ depend only on one Majorana phase. This implies correlations between LNV observables of different flavors in this limit. Our analysis focuses on the approximate relation between $\mee$ and $\mmue$.

\subsection{Unified naming scheme for neutrino masses}

We introduce the following naming scheme for the three neutrino mass eigenstates,
\beq
    \{ \nu_\ell, \ \nu_2, \ \nu_o \}\,,
\eeq
where the $\ell$ stands for {\it ``lightest"} and the $o$ for {\it ``other"}.
This notation utilizes that, independent of the mass ordering, $\nu_2$ is never the lightest mass eigenstate. Any expression can then be written in a simple manner, in terms of these three states, with the following correspondence to the conventional notation for the mass eigenstates:
\begin{equation}\label{eq:convDef}
\tabcolsep 16pt
\begin{tabular}{ll}
    \multicolumn{1}{c}{\rm NO}  &  \multicolumn{1}{c}{\rm IO} \\[4pt]
    $\nu_1 \to \nu_\ell$\,,  &  $\nu_1 \to \nu_o$\,, \\ 
    $\nu_2 \to \nu_2$\,,  &  $\nu_2 \to \nu_2$\,, \\
    $\nu_3 \to \nu_o$\,,  &  $\nu_3\to \nu_\ell$\,.
\end{tabular}
\end{equation}
This convention for the mass eigenstates 
is convenient when one is interested in the behavior for different values of the lightest mass. For other purposes it may be more useful to work with the standard convention. For instance, this choice is useful when we expand in the small parameter $(m_\ell/ m_2)$, while it makes an expansion in the small $s_{13}$ less transparent.

%%%%%
\subsection{Nearly massless lightest neutrino}

Using the above notation for the mass eigenstates, we consider the scenario in which neutrino masses are hierarchical, with the light eigenstate much lighter than the others,
\beq\label{hierarchy}
    m_\ell \ll m_o\, .
\eeq 
We can rewrite Eq.~\eqref{eq:masteree} in the form of an expansion, 
\begin{align}\label{eq:master2}
\malphabeta^2 ={} & m_2^2\, |U_{\alpha 2}|^2\, |U_{\beta 2}|^2 
\nn\\
& \times \Bigg[ 1 +  \sum_{i = o,\ell}\frac{m_i^2}{m_2^2}\frac{|U_{\alpha i}|^2|U_{\beta i}|^2}{|U_{\alpha 2}|^2|U_{\beta 2}|^2} + 
 2\!\!\!\!\!\!\sum_{\substack{(ij)=\\ \{o\ell, 2\ell , 2 o \}}}\!\!\!\!\!\frac{m_i m_j}{m_2^2}  \frac{|U_{\alpha i}U_{\alpha j}U_{\beta i}U_{\beta j}|}{|U_{\alpha 2}|^2 |U_{\beta 2}|^2} \cos\Big(\Phi_{ i j}+ \delta^{\alpha e}_{ij} + \delta^{\beta e}_{ij}
    \Big) \Bigg]\, ,
\end{align}
where we choose to factor out $m_2^2\, |U_{\alpha 2}|^2\, |U_{\beta 2}|^2$, since $m_2$ is nonzero regardless of the ordering.

Consider the limit in which the lightest neutrino mass approaches zero,
\beq
    m_\ell\to 0 \, .
\eeq
In this limit there is only a single Majorana phase, since one phase can be rotated away due to the enhanced symmetry.\footnote{To prove this, following Ref.~\cite{Nir:2001ge}, note that a symmetric rank-2 neutrino Yukawa matrix has 5 real and 5 imaginary parameters (for rank-3, it is 6 real and 6 imaginary). 
The charged lepton Yukawa couplings contain $9+9$ parameters.  The global $U(3)_L\times U(3)_E$ symmetry is still completely broken, allowing to remove $6+12$ parameters, leaving 5 masses, 3 mixing angles, and 2 phases (one Dirac and one Majorana) as physical parameters.}
Thus, only a single cosine term survives in the sum in Eq.~\eqref{eq:master2}. When approaching this limit, for small but nonzero $m_\ell$, two Majorana phases remain physical, but the sensitivity of $m_{\alpha\beta}^2$ to the second phase is suppressed by $m_\ell/m_o$. We refer to the $m_\ell\to 0$ regime as the \textit{single phase} limit.

To make this more concrete, we consider the coefficients of the three cosine terms in Eq.~\eqref{eq:master2}, 
\begin{align}\label{eq:phaseCs}
    C_{o \ell}\,  & = \, \frac{m_o m_\ell}{m_2^2}\frac{|U_{\alpha o}||U_{\beta o}||U_{\alpha \ell}||U_{\beta \ell}|}{|U_{\alpha 2}|^2|U_{\beta 2}|^2} \, , \\ \nn
     C_{2 \ell}\,  & = \, \frac{m_\ell}{m_2}\frac{|U_{\alpha \ell}||U_{\beta \ell}|}{|U_{\alpha 2}||U_{\beta 2}|} \, , \\ \nn
    C_{2 o}\,  & = \, \frac{m_o}{m_2}\frac{|U_{\alpha o}||U_{\beta o}|}{|U_{\alpha 2}||U_{\beta 2}|} \, , \nn  
\end{align}
where the flavor indices, $\alpha$ and $\beta$, in the coefficients $C$ are suppressed.
We can then better define the regime of Eq.~\eqref{hierarchy} by considering the following condition,
\beq\label{eq:singlephase}
    \max\left[C_{2\ell},C_{o \ell}\right] \, < \, 0.1 \, C_{2 o}\, ,
\eeq
where we use an arbitrary choice of $0.1$ as a benchmark for the measure of dominance of the leading coefficient.
The condition in Eq.~\eqref{eq:singlephase} is flavor dependent.
It is known experimentally that the magnitudes of most PMNS matrix elements are ${\cal O}(1)$, and do not play a significant role in determining the hierarchy of the coefficients in Eq.~\eqref{eq:phaseCs}. 
However, for observables involving electrons, the relative smallness of $|U_{e3}| \approx 0.15$ can be important. 

\begin{figure}[tb]
\centerline{
\raisebox{-2pt}{\includegraphics[width=0.345\textwidth]{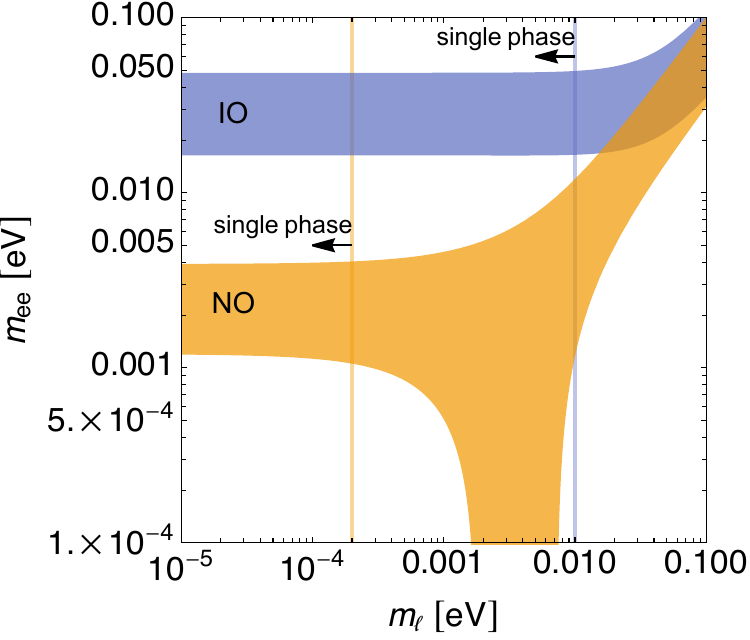}}\label{fig:fig1a_mee}
\includegraphics[width=0.31\textwidth]{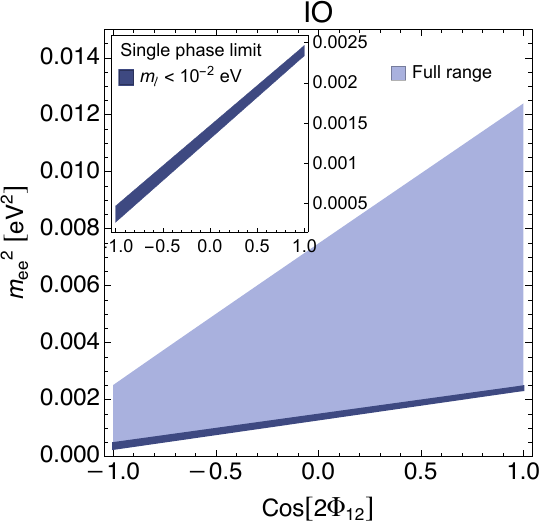}
\includegraphics[width=0.31\textwidth]{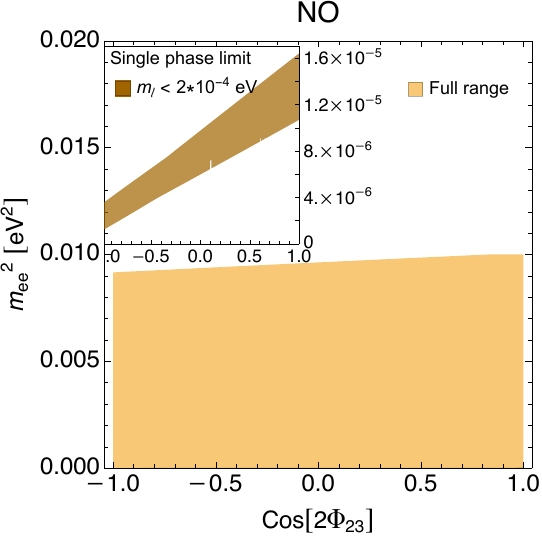}}
\raggedright{\hspace*{3.1cm} (a) \hspace*{4.75cm} (b) \hspace*{4.7cm} (c)}
\caption{The behavior of $\mee^2$ in the single phase limit. (a)~$\mee$ as a function of $m_\ell$, when varying all parameters within $2\sigma$, using the values in Table~\ref{tab:values}. 
The phases $\delta$, $\eta_1$, and $\eta_2$ are unconstrained. The vertical lines show the onset of the single phase limit for IO and NO, as defined in Eq.~(\ref{eq:singlephase}). 
(b)~The correlation of $\mee^2$ and $\cos(2\Phi_{2o})$ in the single phase limit (dark shaded region) vs.\ the correlation in the full range (light shaded region), for IO. 
(c)~The same as in (b) but for NO. In this case the single phase limit correlation cannot be seen in the full range due to the vast range of $\mee^2$ values (see inset plot).}
\label{fig:LinearCos2Phi}
\end{figure} 

The approximate single phase expression is given by
\begin{equation}\label{eq:master3}
	\malphabeta^2 \approx m_2^2\, |U_{\alpha 2}|^2\, |U_{\beta 2}|^2\Bigg[ 1 +  \frac{m_o^2}{m_2^2}\, \frac{|U_{\alpha o}|^2|U_{\beta o}|^2}{|U_{\alpha 2}|^2|U_{\beta 2}|^2} + 2\frac{m_o}{m_2}\, \frac{|U_{\alpha o}||U_{\beta o}|}{|U_{\alpha 2}| |U_{\beta 2}|} \cos\big(2\Phi_{ 2 o}+\delta^{\alpha e}_{o2}+\delta^{\beta e}_{o2}\big) \Bigg] ,
\end{equation}
and hereafter we use the approximately equal sign to denote equalities holding up to a relative correction of $\max\left[C_{2\ell},C_{o \ell}\right]$.
We note that:
\begin{enumerate}
\item Only one Majorana phase, $\Phi_{2o}$, appears in Eq.~\eqref{eq:master3}, regardless of the flavor indices $\alpha$ and $\beta$. This implies that, in principle, assuming the Dirac phase is determined by other means, only one of the six elements of the symmetric matrix \malphabeta is independent in this limit, and can serve to predict all others within the \nuMSM.
\item The approximate relation between \malphabeta and the Majorana phase, $\Phi_{2o}$, is also independent of $m_\ell$ (this was pointed out, for IO, in Ref.~\cite{Simkovic:2012hq}). This implies that, in the single phase limit, the leading Majorana phase information can be extracted from one measurement of \malphabeta, without relying on any knowledge of the absolute mass scale. 
\end{enumerate}

We demonstrate the dependence on a single phase in Fig.~\ref{fig:LinearCos2Phi}, for the case of $\mee$. The single phase limit of Eq.~(\ref{eq:singlephase}) is obtained requiring
\beq
    m_\ell\, \lesssim \, \begin{cases} 10^{-2}\, {\rm eV}\,, & \text{IO} \,, \\  2\times 10^{-4} \, {\rm eV}\,, \quad &  \text{NO} \,. \end{cases}
\eeq
From Fig.~\ref{fig:LinearCos2Phi}(a), we see that in this regime $m_{ee}$ does not depend on the value of $m_\ell$. From panels (b) and (c) (darker regions and insets), we see that a measurement of \mee is a measurement of $\cos(2\Phi_{2o})$.
Outside of the single phase limit (lighter regions), a single Majorana phase cannot determine \mee.  
We emphasize that no prior knowledge on the Dirac phase is assumed in Fig.~\ref{fig:LinearCos2Phi}.

\subsection{Using \texorpdfstring{$\mee$}{} to predict \texorpdfstring{$\mmue$}{}}

\begin{figure}[tb]
\centerline{
\includegraphics[width=0.33\textwidth]{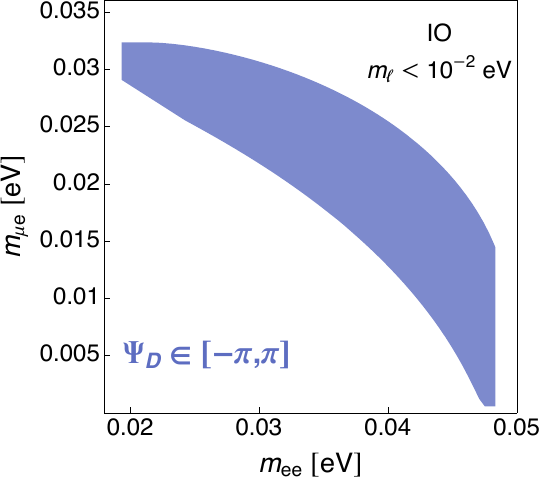}
\includegraphics[width=0.32\textwidth]{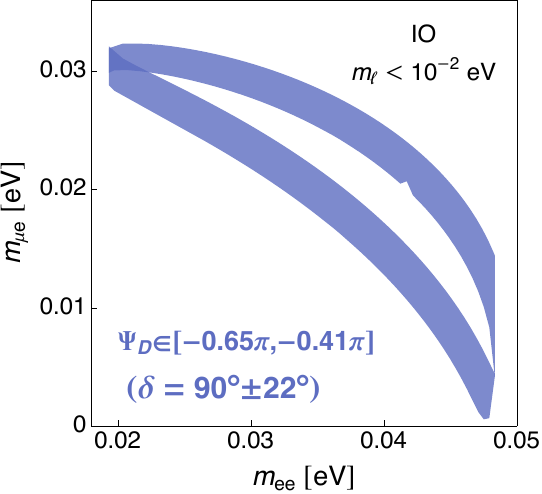}
\includegraphics[width=0.32\textwidth]{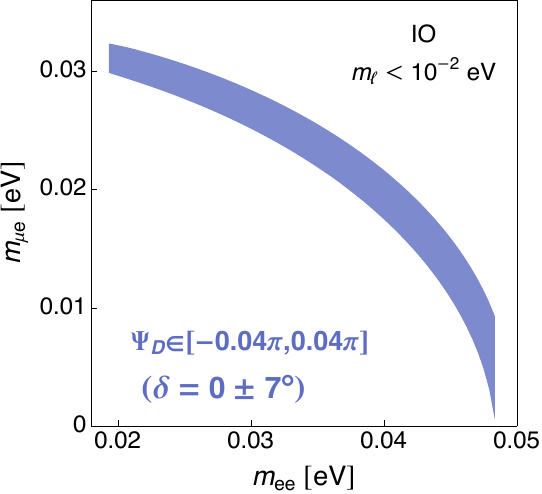}}
\raggedright{\hspace*{3.1cm} (a) \hspace*{4.75cm} (b) \hspace*{4.7cm} (c)}

\vspace*{8pt}
\centerline{
\includegraphics[width=0.32\textwidth]{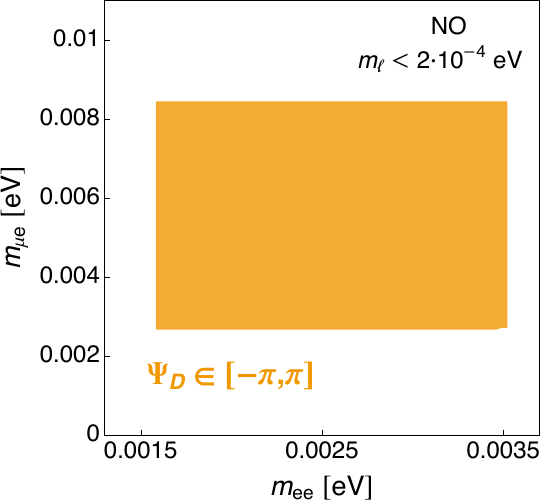}
\includegraphics[width=0.32\textwidth]{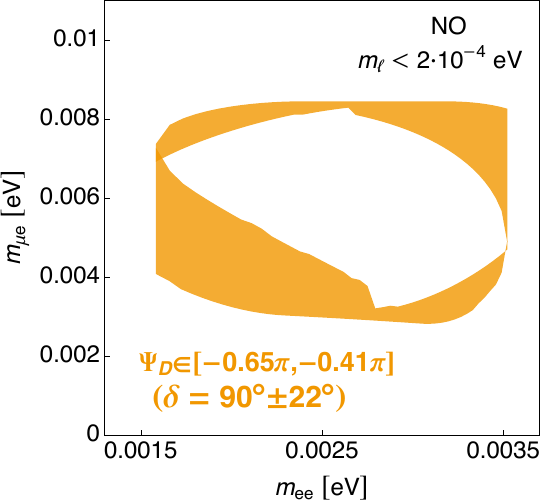}
\includegraphics[width=0.32\textwidth]{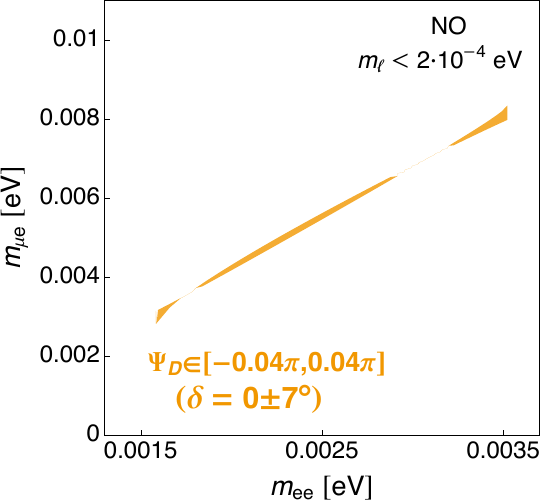}}
\raggedright{\hspace*{3.1cm} (d) \hspace*{4.75cm} (e) \hspace*{4.7cm} (f)}
\caption{The correlation between $\mmue$ and $\mee$ in the single phase limit, under various assumptions on the Dirac phase. The mixing angles and mass squared differences are fixed at their central values shown in Table~\ref{tab:values}. The values for the Dirac phase, $\Psi_D$, are fixed by the given ranges for $\delta$, using the relation of Eq.~\eqref{eq:3phaseRelation}. }
\label{fig:MueMee}
\end{figure} 

In the following we demonstrate the ability to use one element of the \malphabeta matrix to predict another, using the example of \mee and \mmue.
Inverting Eq.~(\ref{eq:master2}), one can see that the value of $\cos2\Phi_{2o}$ is fixed by a measurement of $\mee^2$ in \znbb decays 
up to relative corrections of order 
\beq
\max\left[\frac{m_\ell}{m_2}\frac{|U_{e\ell}|^2}{|U_{e2}|^2},\, \frac{m_\ell}{m_o}\frac{|U_{e\ell}|^2}{|U_{e o}|^2}\right].
\eeq
In particular, if we neglect the subleading corrections in Eq.~(\ref{eq:masteree}), we arrive at the relation
\beq\label{Eq:mee32}
    \mee^2 \approx \sum_i m_i^2\, |U_{ei}|^4 + 2 m_2 m_o\, |U_{e2}|^2|U_{eo}|^2 \cos 2\Phi_{2o} \,.
\eeq

The effective Majorana mass relevant for $\mu^-\to e^+$ conversion, $\mmue^2$, is then related to $\mee^2$, since it is also, to leading order, dependent only on the Majorana phase $\Phi_{2o}$,
\beq\label{eq:mmueapprox}
    \mmue^2 \approx \sum_i m_i^2\, |U_{ei}|^2|U_{\mu i}|^2 
    + 2 m_2 m_o\, \big| U_{e2}U_{eo}U_{\mu 2}U_{\mu o} \big| \cos\big(2\Phi_{2 o}+\delta^{\mu e}_{2 o}\big)\, .
\eeq
We emphasize that the phase $\delta^{\mu e}_{2o}$ contains only Dirac phase information and we assume it would be determined in oscillation experiments. 
More concretely, for the two mass orderings the phase shifts are given by
\beq\label{eq:deltamue}
    \delta^{\mu e}_{2 o} = \left\{\begin{array}{lll} 
    \arg(t_{\mu 2 e 3}) &= \Psi_D \,, & \text{NO} \,, \\*
    \arg(t_{\mu 1 e 2}) &=\displaystyle\arg\left(-\frac{|U_{\mu 2}U_{e2}|}{|U_{\mu 3}U_{e 3}|} -e^{-i\Psi_D}\right) ,\quad  & \text{IO} \,. \end{array} \right. 
\eeq
Figure~\ref{fig:MueMee} demonstrates the correlation between \mmue and \mee in the single phase limit, where, depending on our knowledge of the Dirac phase, a measurement of one can be used in principle to predict the other. 
In the case of inverted ordering, the phase shift $\delta_{2o}^{\mu e}$ in Eq.~(\ref{eq:mmueapprox}) is close to $\pm \pi$, see Eq.~\eqref{eq:deltamue}, and there is a clear (anti-)correlation even without assuming knowledge of the Dirac phase. 
For normal ordering, the value of the Dirac phase strongly affects the correlation.

%%%%
\subsection{Discrete ambiguities}

A measurement of $\malphabeta$ in the single phase limit is a measurement of $\cos2\Phi_{2o}$ (as in Eq.~(\ref{Eq:mee32}) for the case of $\mee$), leaving the four-fold ambiguity in the Majorana phase~\cite{Grossman:1997gd}, $\{\Phi_{2o},\, \pi- \Phi_{2o},\, -\Phi_{2o},\,\Phi_{2o} -\pi\}$, where the phases are defined by convention in the range $[-\pi,\,\pi]$. 
A determination of $\text{sign}(\sin 2\Phi_{2o})$ would reduce the ambiguity to a two-fold, $\{\Phi_{2o},\,\Phi_{2o}-\pi\}$, while resolving the remaining ambiguity would require a measurement sensitive to $\text{sign}(\cos\Phi_{2o})$ or $\text{sign}(\sin\Phi_{2o})$.

In the case of $\mee$ and $\mmue$, while a measurement of $\mee$ would fix the value of $\cos 2\Phi_{2o}$, the different flavor structure of $\mmue$ implies a phase shift in the cosine term compared to $\mee$, as can be seen in Eq.~\eqref{eq:mmueapprox} vs.\ Eq.~\eqref{Eq:mee32}, resulting in a dependence on the product $\sin2\Phi_{2o}\sin\delta^{\mu e}_{2o}$. 
Assuming knowledge of the Dirac phase (magnitude and sign), this implies that a prediction of $\mmue$ from $\mee$ would involve a two-fold discrete ambiguity, corresponding to the unknown sign of $\sin 2\Phi_{2o}$. This ambiguity can be seen in the middle panels of Fig.~\ref{fig:MueMee}, where, for a specific (non zero) value of the Dirac phase, a given value for $\mee$ corresponds to two possible $\mmue$ values.

%%%%%%%%%%%%%%%%%%%%%%%%%%%%%%%%%%%%%%
%%%%%%%%%%%%%%%%%%%%%%%%%%%%%%%%%%%%%%
\section{\texorpdfstring{\mmue}{}, \texorpdfstring{$m_{\mu\mu}$}{} and a no-lose proposition}
\label{sec:nolose}

If the neutrino mass ordering is inverted, upcoming \znbb experiments will determine whether neutrino mass terms violate or conserve lepton number.  However, if the mass ordering is normal, and upper bounds on \znbb improve in the coming decades without discovering a signal, then it becomes a pivotal question to find other means of revealing the nature of neutrino mass. Here we show that a combination of bounds on $m_{ee}$ and $m_{\mu e}$ (or $m_{\mu\mu}$) could, in principle, exclude the \nuMSM also for the case of NO.

\begin{figure*}[t!]
\centerline{
\includegraphics[width=0.45\textwidth]{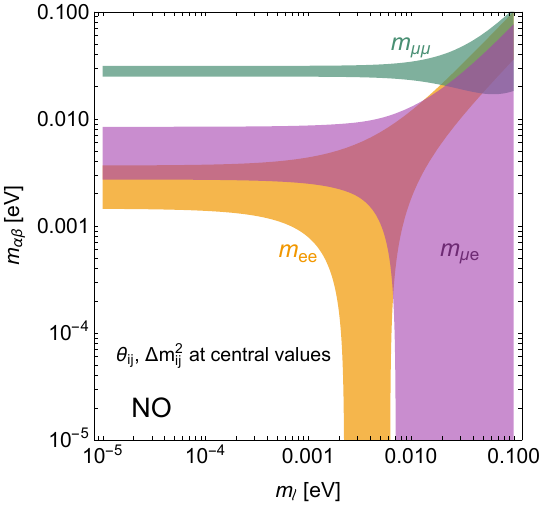}
\hfil
\includegraphics[width=0.45\textwidth]{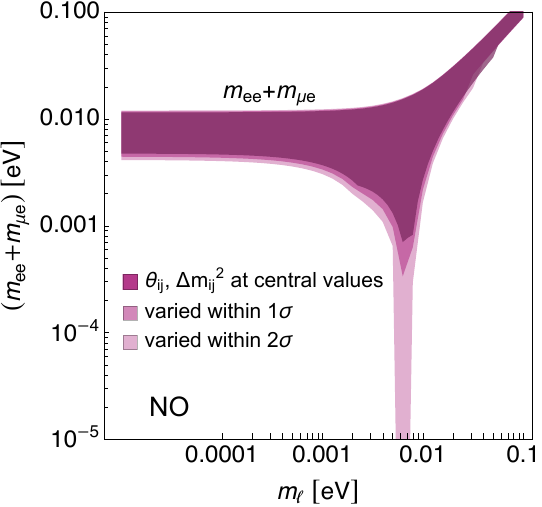}}
\vspace*{4pt}
\raggedright{\hspace*{4.6cm} (a) \hspace*{7.3cm} (b)}
\caption{The allowed ranges of (a) $\mee$, $\mmue$, and $m_{\mu\mu}$ for normal ordering, for the central values of the three mixing angles and two mass-squared differences in Table~\ref{tab:values}. 
In (b) we show the allowed ranges of $\mee+\mmue$ for the central values of these five parameters, and varying them by $1\sigma$ or $2\sigma$.}
\label{fig:magic}
\end{figure*} 

In Fig.~\ref{fig:magic} we plot, for normal ordering, the allowed ranges of \mee, \mmue, and $m_{\mu\mu}$, as functions of the lightest neutrino mass.  
The smallest value of $m_\ell$ for which \mmue can vanish is highly sensitive to the mixing angles.  The left plot shows that taking the central values of the mixing angles in the ``w/o SK-ATM" fit in the PDG~\cite{PDG24} in Table~\ref{tab:values}, the regions of $m_\ell$ for which \mee and \mmue can vanish do not overlap.  
The fact that \mee and \mmue cannot simultaneously vanish means that, at least in principle, one can determine the nature of the neutrino mass, even if \mee is zero.  
While the required \mmue sensitivity is many orders of magnitude beyond that of any proposed experiment for $\mu^-\to e^+$ conversion, this provides, in principle, a no-lose theorem for determining the nature of neutrino mass.
Allowing for the variations of the mixing angles by $2\sigma$ of their present uncertainties, the regions of vanishing \mee and \mmue start to overlap.  
This is shown in the right plot in Fig.~\ref{fig:magic}, where the allowed range for the sum, $\mee+\mmue$, is plotted.
We find,
\beq
    \mee+\mmue > \begin{cases}
        7\times 10^{-4}\, {\rm eV}\,,  &  \text{central values\,,} \\ 
        2\times 10^{-4}\, {\rm eV}\,,  &  \text{within $1\sigma$ ranges}\,,
    \end{cases}
\eeq
where in the second line we varied independently all parameters in Table~\ref{tab:values} within their respective $1\sigma$ ranges. 
Varying them by $2\sigma$, $m_{ee}+m_{\mu e}$ can vanish.
The existence or absence of this overlap region will also be impacted by future constraints on the $CP$ violating phase $\delta$, and on the lightest neutrino mass, $m_\ell$.
In any case, this adds to the motivations to determine the mixing parameters more precisely.  

\begin{figure*}[t!]
\centerline{
\includegraphics[width=.45\textwidth]{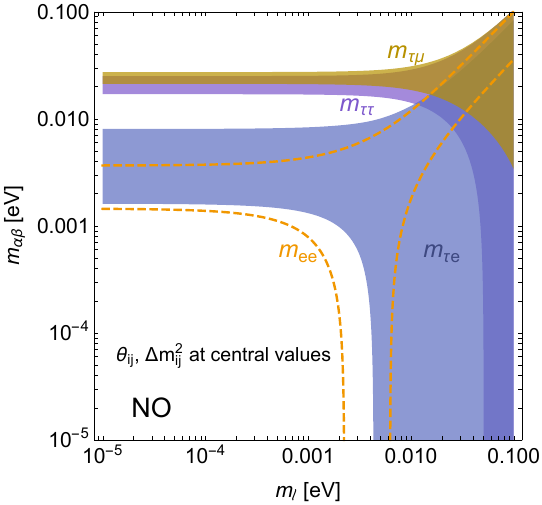}
}
\vspace*{-8pt}
\caption{The allowed ranges of $m_{\tau e}$, $m_{\tau \mu}$, and $m_{\tau \tau}$, compared to $\mee$ (dashed contour), as a function of the lightest neutrino mass in the case of normal ordering and for the central values of the three mixing angles and two mass-squared differences in Table~\ref{tab:values}.}
\label{fig:tau}
\end{figure*}

For completeness, Fig.~\ref{fig:magic}\,(a) also shows the allowed range for $m_{\mu\mu}$, and Fig.~\ref{fig:tau} shows the corresponding plot for $m_{\tau e}$, $m_{\tau\mu}$, and $m_{\tau\tau}$. For normal ordering, neither $m_{\tau\mu}$ nor $m_{\tau\tau}$ can vanish simultaneously with $m_{ee}$. 
The prospects for experimentally probing these elements of the $m_{\alpha\beta}$ matrix are, however, even more 
remote than for $\mmue$.

%%%%%%%%%%%%%%%%%%%%%%%%%%%%%%%
%%%%%%%%%%%%%%%%%%%%%%%%%%%%%%%
\section{Conclusion}
\label{sec:conclusion}

The nature of neutrino masses is a fundamental open question. 
If neutrinos are Majorana particles, then their mixing involve two yet unconstrained $CP$ violating phases, which only affect lepton number violating processes.
Measuring \znbb would constrain one linear combination. 
Our analysis of $\Delta L=2$ processes within the \nuMSM leads to the following conclusions:
\begin{enumerate}
    \item The sensitivity of $\malphabeta^2$ to a specific Majorana phase scales with the product of the corresponding neutrino masses, that is, for a small but nonzero lightest mass, $m_\ell$, the sensitivity to the second Majorana phase is reduced linearly.

    \item In the regime of a very small
    $m_{\ell}$, one Majorana phase, $\Phi_{2o}$, is dominant, and its relation to $\malphabeta$ is approximately independent of $m_\ell$ itself. The dominance of a single Majorana phase is common to all entries of \malphabeta, implying that at leading order only one of the six elements is independent in this limit.

    \item If future progress allows reaching \nuMSM sensitivity for \mmue, the Majorana nature of neutrinos could be ruled out by the non-observation of both \mee and \mmue. 
    (The same can also be achieved if $m_{\mu\mu}$ can be probed.)
    The current central values of the mass-squared differences and PMNS mixing angles yield no overlap for the regions where \mee and \mmue both vanish. 
    This provides additional motivation for precision measurements of mixing parameters, as well as for further exploration of ways to probe \mmue. 
\end{enumerate}

Although our results pertain to observables beyond current sensitivity,  
furthering our understanding of lepton number violating processes 
is essential to characterize the basic nature of the lepton sector.
Upcoming results from various experiments (cosmological, \znbb, $m_{\nu_e}$, oscillation) will contribute to this developing understanding.

\acknowledgments

ZL thanks Hitoshi Murayama and Yury Kolomensky for related conversations many years ago.
The work of YG is supported in part by the NSF grant PHY1316222.
The work of ZL is supported in part by the Office of High Energy Physics of the U.S.\ Department of Energy under contract DE-AC02-05CH11231. 
The research of SG is supported in part by the U.S. Department of Energy grant number DE-SC0010107.
AD is supported by the Women's postdoctoral career development award of the Weizmann Institute of Science.

% If not commented out, lists all entries in bib file
%\nocite*
\bibliography{refs}

\end{document}